\documentclass[english,12pt,nofootinbib,onecolumn,tightenlines,a4paper,notitlepage]{revtex4-1}  

\usepackage{amsmath}
\usepackage{amssymb}
\usepackage{amsthm}
\usepackage{bbm}

\usepackage{hyperref}
\usepackage{cleveref}

\theoremstyle{definition}
\newtheorem{definition}{Definition} 

\theoremstyle{plain}
\newcounter{tlc}

\newtheorem{lemma}[tlc]{Lemma}

\crefname{theorem}{theorem}{theorems}
\crefname{lemma}{lemma}{lemmata}
\crefname{corollary}{corollary}{corollaries}
\newtheorem*{theorem*}{Theorem}

\usepackage{enumerate}

\usepackage{graphicx}
\usepackage{epstopdf}
\usepackage[font={small,it},labelfont={bf,sf},margin=0.5cm]{caption}

\newcommand{\Or}[1]{\mathrm{O}\!\left(#1\right)}
\newcommand{\SO}[1]{\mathrm{SO}\!\left(#1\right)}
\newcommand{\SU}[1]{\mathrm{SU}\!\left(#1\right)}
\newcommand{\Sp}[1]{\mathrm{Sp}\!\left(#1\right)}

\newcommand{\Un}[1]{\mathrm{U}\!\left(#1\right)}
\newcommand{\quat}{\mathbb{H}}

\newcommand{\bra}[1]{\langle #1|}
\newcommand{\ket}[1]{|#1\rangle}
\newcommand{\braket}[2]{\langle #1|#2\rangle}
\newcommand{\ketbra}[2]{|#1\rangle\!\langle#2|}
\newcommand{\chainketbra}[2]{#1\rangle\!\langle#2|}
\newcommand{\expt}[1]{\langle #1 \rangle}

\newcommand{\id}{\mathbbm{1}}
\newcommand{\herm}{^{\dag}} 
\newcommand{\trans}{^\mathrm{T}}

\DeclareMathOperator{\tr}{tr}

\DeclareMathOperator{\diag}{diag}

\newcommand{\Real}{\Re\mathrm{e}}
\newcommand{\Img}{\Im\mathrm{m}}

\newcommand{\inlineheading}[1]{\textbf{{#1---}}}

\newcommand{\covec}[1]{\reflectbox{\ensuremath{\vec{\reflectbox{\ensuremath{#1}}}}}}

\newcommand{\spek}[1]{\raisebox{-0.25em}{\includegraphics[width=1.1em]{./spek#1}}}

\begin{document} 

%% TITLE
\title{Interferometric computation beyond quantum theory}

%% AUTHOR
\author{Andrew J.\ P.\ \surname{Garner}}
\email{ajpgarner@nus.edu.sg}
\affiliation{Centre for Quantum Technologies, National University of Singapore,
3 Science Drive 2, 117543, Singapore}

%% DATE
\date{March 16, 2018}

%% ABSTRACT
\begin{abstract}
There are quantum solutions for computational problems that make use of \mbox{{\em interference}} at some stage in the algorithm.
These stages can be mapped into the physical setting of a single particle travelling through a many-armed \mbox{interferometer}.
There has been recent foundational interest in theories beyond quantum theory.
Here, we present a generalized formulation of computation in the context of a many-armed interferometer, 
 and explore how theories can differ from quantum theory and still perform distributed calculations in this set-up.
We shall see that {\em quaternionic quantum theory} proves a suitable candidate, whereas  {\em box-world} does not.
We also find that a classical hidden variable model first presented by Spekkens\footnote{Phys. Rev. A, 75:3:32100, 2007} can 
 also be used for this type of computation due to the epistemic restriction placed on the hidden variable.
\end{abstract}

%% RENDER TITLE
\maketitle

%%
%% INTRO
\section{Introduction}
There are algorithms available for quantum computers that perform particular tasks faster than any known deterministic classical algorithm~\cite{Deutsch85,DeutschJ92,Simon94,Grover96,Shor97,CleveEMM98,Lloyd99}.
It is argued by~\citet{CleveEMM98} that the quantum efficiency arises from the ability
 of quantum computers to usefully leverage interference within their algorithms: a process that is not possible with a classical computer.
That is, quantum algorithms
 incorporate stages where one prepares a superposition of bit-strings, 
 enters this into some device (typically determined by the input to the problem),
 and then {\em interferes} the output from this device in order to extract information useful for solving the computational task.

There has been much foundational interest in operational theories that go beyond quantum theory, perhaps incorporating quantum theory as a limiting case, or even replacing it in yet-unencountered regimes.
Often, one considers a set of fundamental principles ({\em axioms}) and uses these to derive quantum theory~(see e.g.\ \cite{Zeilinger99,Hardy01,Fuchs01,MasanesM11,ChiribellaDAP12,BarnumMU14}).
By testing these foundational principles one can validate quantum theory as a good description of reality; by relaxing them one can create {\em foil theories} (alternatives) that exhibit contrasting behavior to quantum theory, which might be observable in experimental tests.

In particular, one class of post-quantum theories known as {\em box-world} concerns itself with theories that are more non-local than quantum theory 
\cite{ClauserHSH69,Cirelson80,Tsirelson93,PopescuR94,Popescu14}.
If such non-local states existed, it has been shown that they would make certain tasks involving distributed computing trivial~\cite{vanDam00,ClevevDNT13}.
Moreover, recent studies have explored the possibility of computation in post-quantum theories~\cite{FernandezS03,LeeB15,LeeH16,LeeS16,LeeS16b}, 
 particularly showing that non-trivial interference in a theory may be used as a resource for computation beyond the known classical limit~\cite{LeeS16b}.

\begin{figure}[tbh]
\begin{center}
\includegraphics[width=0.65\textwidth]{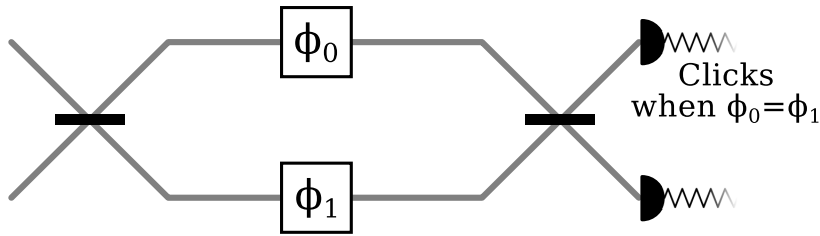}
\caption[Mach--Zehnder interferometer]{
\textbf{Mach--Zehnder interferometer.}
A photon incident from the upper branch is put into a superposition of both spatially disjoint {\em branches} by the first beamsplitter.
Phases $\phi_0$ and $\phi_1$ are induced on each arm.
The branches recombine at the second beamsplitter, and if $\phi_0=\phi_1$, then the photon finishes in the upper branch with certainty.
If $\,\left|\phi_0 - \phi_1\right| = \pi$, then the photon finishes in the lower branch with certainty.
}
\label{fig:MZI}
\end{center}
\end{figure}

Here, we shall consider the implementation of computation specifically in the context of {\em branching interferometers}---%
 devices such as the {\em Mach--Zehnder interferometer} (\cref{fig:MZI}) that have set of spatially disjoint arms traversed by a single particle.
Building on recent work that extends the concept of interference beyond quantum theory~\cite{Sorkin94,UdudecBE10,Ududec12,GarnerDNMV13,DahlstenGV14,Garner15DPhil}, 
 we consider the case where the traversing particle has a spatial ``which branch?'' degree of freedom described by a post-quantum theory.
We shall probe whether the interferometric devices described by these theories allow for the behaviour required for computation.

%%
%% FIGURE: MZI

In particular, we will focus on the interferometric implementation of the Deutsch--Jozsa algorithm~\cite{Deutsch85,DeutschJ92,CleveEMM98} and Grover's algorithm~\cite{Grover96,Lloyd99}.
We will discuss how to represent this type of algorithm interferometrically in a generalized setting.
We will then show that this cannot be implemented by {\em gbits} in box-world, but can be extended into quaternionic quantum theory.
Finally, we will consider the case of Spekkens' toy model~\cite{Spekkens07,Spekkens14}, a classical hidden variable theory, which is made to exhibit certain aspects of quantum behavior by imposing a restriction on the allowed states of knowledge of the system. 
In this case, we find that it is explicitly the epistemic restriction that makes interferometric computation possible,
 even though non-trivial dynamics are also possible on the states of the underlying hidden variable.

%%
%% TECH INTRO
\section{Introduction to post-quantum theories}
\noindent
% INTRO: GPTS
\inlineheading{The GPT framework.}
We will find it convenient to adopt the framework of generalized probabilistic theories (GPTs) as presented for instance in~\cite{Hardy01,Barrett07,MasanesM11}.
This framework amounts to taking the operational view that a system can be described in terms of the measurements made on it, and the outcomes associated with any given measurement.
The states of the system are then a collection of probability distributions over these measurements.
It is assumed that by knowing the statistics of some finite number of different measurements (an {\em informationally complete set}), 
 one can infer the probabilities of the outcomes of {\em any} measurement made on the system.
These inferred probabilities are known as the {\em state} of the system, which can thus be written as a real-valued vector $\vec{s}$.
Likewise, there is a natural dual space of measurement outcome {\em effects}, which are functionals that act on states to give the probability associated with any particular measurement outcome.
An effect is written as a real-valued (co-)vector $\covec{e}$, such that for some outcome $i$ associated with effect $\covec{e_i}$, the probability that this outcome occurs for a state $\vec{s}$ is given by the inner product $P\left(i |\vec{s}\right) = \covec{e_i}\cdot\vec{s}$.

For example, consider a (classical) coin which has either ``heads'' or ``tails'' facing upwards.
The state of the coin may be written as $\vec{s}=\left[ P\!\left(\mathrm{heads}\right),\, P\!\left(\mathrm{tails}\right) \right]\trans$, and the effects associated with heads and tails are $\covec{e_\mathrm{H}} = \left(1, 0\right)$ and $\covec{e_\mathrm{T}} = \left(0, 1\right)$ respectively.
By inspection, $\covec{e_\mathrm{H}}\cdot\vec{s} = P\!\left(\mathrm{heads}\right)$. 
Following a fair toss, the state would be written $\vec{s}=\left(\frac{1}{2}, \frac{1}{2}\right)$ and the probability of being in heads (or tails) is obviously given by $\frac{1}{2}$.

A more involved example would be to describe the state of the elementary two-level quantum system: the qubit.
By knowing the outcome probabilities of measurements associated with the Pauli matrices ($\sigma_x$, $\sigma_y$ and $\sigma_z$), one can fully recreate the state (i.e.\ density matrix in the quantum formalism).
Thus, within the GPT framework, one can express the qubit state as a real positive vector:
\begin{equation}
\label{eq:3in2out}
\vec{s} = \left(\begin{array}{c}
P\!\left(X=+1\right) \\
P\!\left(X=-1\right) \\
\hline
P\!\left(Y=+1\right) \\
P\!\left(Y=-1\right) \\
\hline
P\!\left(Z=+1\right) \\
P\!\left(Z=-1\right) \\
\end{array}\right).
\end{equation}

%%
%% FIGURE: Qubit & Gbit state-space
\begin{figure}[hbt]
\begin{center}
\includegraphics[width=0.75\textwidth]{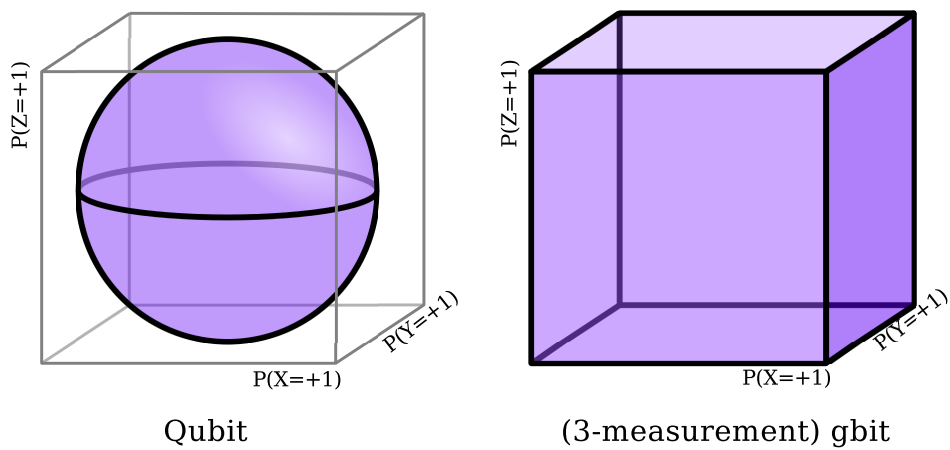}
\caption[Qubit and gbit state-spaces]{
\textbf{Qubit and gbit state-spaces.}
LHS shows the allowed states a qubit can take.
RHS shows the allowed states a (three-measurement) gbit can take.
The uncertainty principle prevents the qubit from taking arbitrary states.
}
\label{fig:SS}
\end{center}
\end{figure}

In quantum theory, there are restrictions on the probability distributions for the measurements within a state.
For instance, a state where $P\!\left(Z=+1\right) = 1$ and $P\!\left(X=+1\right)=1$ is disallowed because it violates the uncertainty principle.
Because the set of eigenvectors of the Pauli matrices form a mutually-unbiased basis, if any Pauli matrix measurement has an outcome which occurs with certainty (probability~$1$), then the outcomes for all other Pauli matrix measurements occur randomly according to a uniform distribution.
The uncertainty restriction is very much a property of quantum theory, following from the Cauchy-Schwarz inequality (see e.g.~\cite{NielsenC00}) as applied to a Hilbert space.
Geometrically, 
 the constraints on allowed statistics imposed by the uncertainty relation (specifically as formulated by Schr\"odinger~\cite{Schrodinger30,Robertson29})
 correspond to restricting the valid states of the qubit to a sphere (see \cref{app:Uncertainty}).
This is illustrated on the LHS of \cref{fig:SS}.

It is possible to conceive of alternative theories where this restriction does not hold.
{\em Gbits}, which form the single-system state-spaces in {\em box-world}, admit any valid probability distribution (subject to the requirement that all measurements normalize to the same value).
A gbit composed of three binary measurements could be written in exactly the same form as the quantum bit (\cref{eq:3in2out}), but without the uncertainty principle, its state-space is given by a cube, as illustrated on the RHS of \cref{fig:SS}.

Finally, we remark that transformations in the GPT framework are linear maps between allowed states.
We shall explicitly consider normalization-preserving transformations.
For classical theories (where there exists a single informationally complete measurement), the normalization-preserving transformations are exactly the set of stochastic matrices acting on a probability vector, whereas in general they have a slightly more complicated form (see {\em Theorem 7} in \cite{Barrett07}).
In all cases, they are real-valued matrices.
Reversible transformations take the state-space to itself and are thus the automorphism group of the state-space (or some sub-group thereof if there are other restrictions imposed on the theory).
For example, single-qubit reversible transformations are given by $\SO{3}\subset \Or{3}$, whereas gbit reversible transformations correspond to the finite group formed by relabelling measurements and measurement outcomes (e.g.\ for three binary measurements, this gives the octahedral group of cubic symmetries).

% INTRO: SPEKKENS
\inlineheading{Spekkens' toy theory.}
Spekkens has presented a classical hidden-variable toy theory that replicates many of the features of quantum theory~\cite{Spekkens07,Spekkens14}, through the introduction of an {\em epistemic restriction} that restricts states to only contain half the amount of knowledge required to fully determine the hidden variable.
The elementary system in this model consists of four classical possibilities (i.e.\ two bits), of which only one bit of information can be known.
We can pictorially represent the four ontic possibilities as $\spek{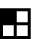}$, $\spek{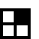}$, $\spek{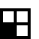}$ and $\spek{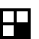}$.

The valid single-bit epistemic states thus consist of two ontic possibilities, and can be divided into three groups of mutually-exclusive pairs: $\big\{\spek{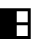}, \; \spek{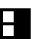}\big\}$,  $\big\{\spek{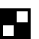}, \; \spek{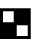}\big\}$  and $\big\{\spek{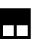}, \; \spek{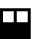}\big\}$, which by analogy with quantum theory we label as the $\pm1$ outcomes of $X$, $Y$ and $Z$ respectively.
These pairs of states are naturally associated with measurements that ask which of the two options is consistent with the underlying ontic state.
Following any measurement, the ontic state is moved to one of the supported possibilities with equal probability (preventing the violation of the epistemic principle by making two different measurements in a row).
From this, it follows that if a measurement is made on an epistemic state which itself is not one of the outcomes of the measurement (e.g.\ measurement $\spek{13}$ vs.\ $\spek{24}$ made on the state $\spek{12}$), then either outcome should occur with equal probability.

%%
%% FIGURE: Spekkens state-space
\begin{figure}[hbt]
\begin{center}
\includegraphics[width=0.75\textwidth]{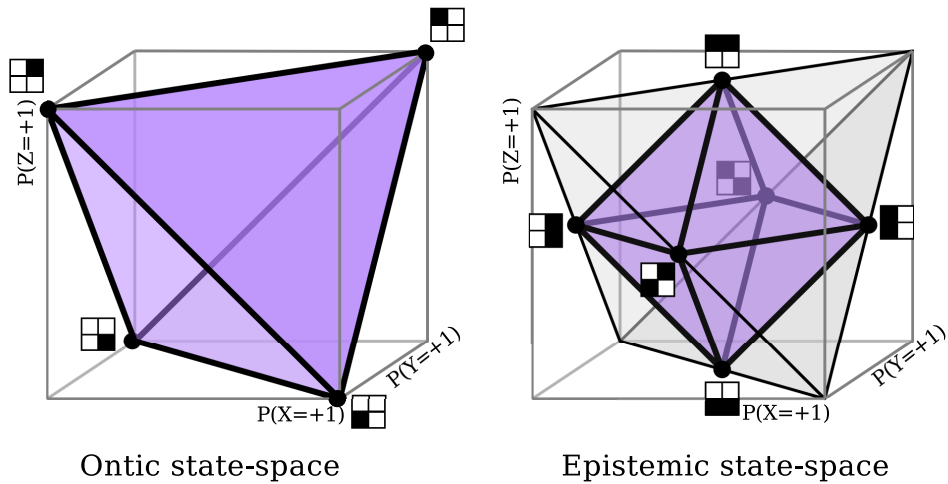}
\caption[State-spaces of Spekkens' toy theory and the underlying hidden variable]{
\textbf{State-spaces of Spekkens' toy theory and the underlying hidden variable.}
LHS shows the {\em ontic} tetrahedron of states that the hidden variable can take.
RHS shows the {\em epistemic} octahedron of states of Spekkens' toy model that satisfy the epistemic restriction.
}
\label{fig:ConvexSpek}
\end{center}
\end{figure}

\enlargethispage{\baselineskip}
Although this theory not {\em a priori} part of the GPT framework, it can be naturally extended by taking the convex combination of the statistics of these states~\cite{GarnerDNMV13} (cf.\ other treatments in \cite{vanEnk07,JanottaL13}).
The states may then be written in terms of the $X$, $Y$ and $Z$ statistics, and represented by a vector in the form of \cref{eq:3in2out}, making it naturally comparable with qubits and three-measurement gbits (sharing the same effects for these measurements).
The set of allowed states can be plotted on Cartesian axes, as shown in \cref{fig:ConvexSpek}.
The statistics spanned without an epistemic restriction form a tetrahedron, whereas with the epistemic restriction, one has an octahedron (identical to the convex hull of one-bit stabilizer states~\cite{Pusey12}).
Because transformations in Spekkens' toy theory correspond to permutations on the hidden variable, for a single bit they are given by the automorphism group of the tetrahedron $S_4$; not all of the symmetries of the octahedron are allowed.

% INTRO: PHASE
\section{Introduction to generalized phase and interference}
In this paper, we shall consider phase as presented in~\cite{GarnerDNMV13,DahlstenGV14,Garner15DPhil} (cf.\ the extension of Sorkin's hierarchical family of interference patterns~\cite{Sorkin94,UdudecBE10,Ududec12}).
Phase between possibilities is one of the most remarkable features that distinguishes quantum theory from classical probability theory, when it comes to describing a degree of freedom: 
Much as knowing the amplitude of a classical wave is not sufficient to predict how two waves interfere, simply knowing the probability associated with a particular possibility is not in general sufficient to predict the behavior of a quantum system.
For example, consider a binary degree of freedom ($0$ vs.\ $1$), described by a qubit (that is, the two ``possibilities'' correspond to outcomes $\ket{0}$ or $\ket{1}$ in the computational basis).
The quantum states $\ket{+}= \frac{1}{\sqrt{2}}\left(\ket{0}+\ket{1}\right)$ and $\ket{-}= \frac{1}{\sqrt{2}}\left(\ket{0}-\ket{1}\right)$, when measured in the computational basis, are equally likely to give either outcome.
However, $\ket{+}$ and $\ket{-}$ differ in their phases between the two possibilities $\ket{0}$ and $\ket{1}$.
Thus, after a Hadamard gate $H=\frac{1}{\sqrt{2}}\big(\begin{smallmatrix}1&1\\1&-1\end{smallmatrix}\big)$ is applied to the two states: $H\ket{+} = \ket{0}$ and $H\ket{-}=\ket{1}$, the two states would give opposite answers in the computational basis with certainty.

For quantum theory in general, for some basis $\{\ket{\xi_j}\}$, all states of the form $\ket{\psi} = \sum_j a_j e^{i\phi_j}\ket{\xi_j}$ for $a_j, \phi_j \in \Re$ with fixed values of $\{a_j\}$ will have the same statistics with respect to the $\{\ket{\xi_j}\}$ measurement for all values of $\{\phi_j\}$.
Moreover, any transformation of the form $U_\xi = \sum_j e^{i\alpha_j} \ketbra{\xi_j}{\xi_j}$ will not alter the $\{\ket{\xi_j}\}$ statistics.
Transformations of this kind are known as {\em phase operations}.
In quantum theory, it is as pertinent to consider the set of allowed unitary matrices of this form as it is the set of allowed values of $\{\phi_i\}$ in $\ket{\psi}$;
 one can generate a state with arbitrary $\{\phi_i\}$ values from some reference state, and free choice of the operation $U_\xi$.

This concept can be extended to the GPT framework as follows~\cite{GarnerDNMV13}:
\begin{definition}[Phase operation]
\label{def:phase}
A transformation $T$ is a {\em phase operation} associated with a measurement $M$ if it never alters the statistics associated with $M$, such that:
\begin{equation}
\covec{e_i}\cdot \vec{s} = \covec{e_i}\cdot T \vec{s}
\end{equation}
for every effect $\covec{e_i}$ associated with $M$, and every state $\vec{s}$ admitted in the theory.
In the case of the reversible transformations, the phase restriction induces a sub-group of transformations which we refer to as the {\em phase group}.
\end{definition}

We shall be interested in particles traversing through {\em branching interferometers} (e.g.\ the Mach--Zehnder interferometer in \cref{fig:MZI}). 
Particularly, one measurement of the GPT state (say, $Z$) distinguishes which {\em branch} (spatially disjoint arm) of the interferometer that the traversing particle is in.
That is, each branch $i$ is associated with a different effect $\covec{z_i}$ of this measurement $Z$, 
 such that if the traversing particle (represented by state $\vec{s}$) is definitely passing through branch $i$, then $\covec{z_i}\cdot\vec{s} = 1$.
For the interferometers we shall consider, these possibilities are mutually exclusive: 
 for all states $\vec{s}$,  $\covec{z_i}\cdot\vec{s} = 1$ implies $\covec{z_j}\cdot\vec{s} = 0$ for all other branches $j\neq i$.
Naturally, there are states where the particle is not in any particular branch with certainty.
For any GPT other than classical theory, the measurements other than the ``which branch?''\ measurement $Z$ encode the ``phase'' information of the state.
If we imagine some physical element that might be added to a branch that does not cause the particle to jump between branches (such as a piece of glass altering the optical path-length), then the operation on the state naturally corresponds to a {\em phase operation} with respect to $Z$. 

However, we need an operational method of identifying whether a particular transformation can be implemented on a particular branch.
It is not {\em a priori} obvious if every transformation in the phase group can be induced by an action taken locally on one particular branch.
In quantum theory, one might by convention associate
\begin{equation}
\label{eq:quantumBL}
U_i = e^{i\Phi}\left( e^{i\phi_i} \ketbra{i}{i} + \sum_{j\neq i}^N \ketbra{j}{j}\right)
\end{equation}
with an action taken on branch $i$. 
The set of all such $\phi_i$ in one branch determines a subgroup of the $\bigoplus^N\!U(1)$ phase group.
This form guarantees that an action taken on, say, branch $1$ does not induce a phase difference between remote branches $2$ and $3$; only for a two-branch system are the two groups are identical, due to the global $\Un{1}$ phase freedom.
As the above form does not obviously generalize beyond Hilbert spaces, we will use the following operational concept to identify the local subgroups in general theories~\cite{DahlstenGV14}:
\begin{definition}[Branch locality]
\label{def:BL}
A transformation $T_i$ can be localized to branch $i$ if it does not alter any states that have no probability of being in branch $i$.
For position measurement effect $\covec{z_i}$ associated with branch $i$, where $\vec{s_{\lnot i}}$ is an arbitrary state with no support on branch~$i$:
\begin{equation}
\covec{e}\cdot\vec{s_{\lnot i}} = \covec{e} \cdot T_i \vec{s_{\lnot i}} \quad
 \forall\covec{e},\, \vec{s_{\lnot i}} \in \{\vec{s} \,|\, \covec{z_i}\!\cdot\!\vec{s}\!=\!0\}
\qquad \leftrightarrow \qquad
T_i \vec{s_{\lnot i}} = \vec{s_{\lnot i}}.
\end{equation}
\end{definition}
Any $T$ that satisfies this for branch $i$ may be said to be {\em localizable} to branch $i$; though this does not imply that $T$ can {\em only} be localized to $i$;
for example, the identity element $\id$ satisfies branch locality for all branches in every theory.
It is shown in \cref{app:BLQT} that \cref{def:BL} implies the conventional quantum form written in \cref{eq:quantumBL}.

This is a {\em locality} restriction, because it captures an intrinsic quality of something being ``over here'' instead of ``over there'' (cf.\ ``non-locality'' in the context of CHSH-Bell violation~\cite{ClauserHSH69,Cirelson80,Tsirelson93,PopescuR94}, which concerns itself with the impossibility of quantum {\em local realism}).
Without branch locality, one would find it conceptually very difficult to ever isolate an experiment from the rest of the universe: every action taken {\em everywhere} would have to be taken into account in order to predict the outcomes of an experiment.
Thus, we will axiomatically impose branch locality as a reasonable restriction on the behavior of any spatially-disjoint interferometer.

%%
%% SECTION: D-J beyond QT
\section{Interferometric computation: constant vs.\ balanced}
\noindent
\inlineheading{Deutsch--Jozsa algorithm.}
Suppose there is some function $f: \{0,1\}^{\otimes n}\to \{0,1\}$, which is guaranteed either to be {\em constant} such that $f(x)$ for some bit-string $x\in\mathcal{B}^n = \big\{\{0,1\}^{\otimes n}\big\}$ evaluates to the same value for all choices of input, or {\em balanced} such that $f\left(x\right)=1$ for exactly half of the inputs (and $0$ for the other half).
Classically one might need to test $f$ with up to $2^{n-1} + 1$ inputs before it could be verified to be constant (a balanced function could be discovered after $2$ inputs, but might look constant for the first $2^{n-1}$ tests).
However, by inputting a quantum bit-string into the device, the Deutsch--Jozsa algorithm~\cite{Deutsch85,DeutschJ92,CleveEMM98} can verify whether the function is constant or balanced in just one query.
The efficient quantum encoding~\cite{CleveEMM98} involves $n+1$ qubits, and a ``quantum oracle'' which performs an $f$-controlled-NOT gate that evaluates $f$ on the first $n$ qubits and applies a NOT gate on the final qubit if $f(x)=1$.
By preparing all of the qubits in the $\ket{+}=\frac{1}{\sqrt{2}}\left(\ket{0}+\ket{1}\right)$ state, the final qubit remains as $\ket{+}$ if the function is constant and goes to $\ket{-}=\frac{1}{\sqrt{2}}\left(\ket{0}-\ket{1}\right)$ if it is balanced.

There is also a simple single-photon interferometry experiment that solves the fixed vs.\ balanced problem using one large degree of freedom.
Suppose each path encodes a different bit-string ($x\in \mathcal{B}^n$). 
A beamsplitter (or series of beamsplitters) can be used to prepare a coherent superposition of all possible bitstrings, written as 
$\left(\frac{1}{\sqrt{2}}\right)^{\!n} \sum_{x\in\mathcal{B}^n} \ket{x}$.
The function $f$ can then be encoded as a collection of phases induced on branch $x$ such that if $f\!(x)=0$ then a phase shift of $0$ is applied to branch $\ket{x}$, and if $f\!(x)=1$ then a phase shift of $\pi$ is applied.
Thus, the unitary phase operation $\sum_{x\in\mathcal{B}^n} e^{i \pi f(x)} \ketbra{x}{x}$ is applied over all $2^n$ branches.
Finally, a set of beamsplitters is used to execute the operation $H^{\otimes n}$.
If $f$ is constant, then the photon will always be found in branch $0$, and if it is balanced the photon will never be found in branch $0$.
By placing a detector in this branch, it is possible to tell if $f$ is balanced or constant in a single run of the experiment.
The single-bit version of this protocol is exactly implemented by a Mach--Zehnder interferometer (\cref{fig:MZI}).

Such an implementation is arguably not as efficient as the Deutsch--Jozsa algorithm, as it requires $2^n$ branches (and hence $2^n$ optical components), but does not require any entangling gates.
However, when treated as an oracle problem, this solution only requires one oracle query: i.e.\ can be performed by passing just one single photon through the system, and so if transmission of data to and from the oracle is the expensive part of the task (say because half of the experiment is very far away, or the oracle charges per access~\cite{Lloyd99}), then this could still be viewed as an efficient solution with respect to oracle calls.

For certain functions, this interferometric approach may still be as good as possible if one also takes into account the complexity of the oracle query itself.
Suppose one randomly picks a function $f:\mathcal{B}^n\to\mathcal{B}$, such that the inputs $x\in\mathcal{B}^n$ that evaluate to $1$ are randomly distributed (i.e.\ do not follow some compressible pattern such as ``the answer is $1$ if the first bit is $1$'').
Specifying this function would be akin to writing down a bit-string of length $2^n$, 
 and the circuitry required to implement it as an oracle would be correspondingly complex.
In the worst case, the oracle would have to distinguish all $2^n$ inputs to decide the answer,
 corresponding to a look-up table with $2^n$ elements.
This approach maps naturally to the interferometric set-up presented here:
 the index of the interferometer's many branches correspond to the value to be looked up, 
 and the presence or absence of a phase plate encodes the stored value in the table.

In this article, we shall focus on the interferometric algorithm, as it can be considered purely in the context of phase and interferometry, in a way that naturally generalizes beyond quantum theory, and allows for interesting exploration of post-quantum dynamics.
%

% GPT D-J
\inlineheading{Interferometric computation in generalized probabilistic theories.}
Let us now address the interferometric computation of the constant vs.\ balanced problem in generalized probabilistic theories.
We will make the assumption that all theories we consider are {\em operational} such that there is a method of preparing any state in the state-space, and that all statistics required to specify the state of the system can be extracted by some kind of measurement (if the experiment is repeated enough times).
This broadly amounts to abstracting away the initial and final beamsplitters (as well as the measurement device) and instead focusing our attention on the central region where the branches are spatially disjoint.

This central region corresponds to an {\em interferometric oracle query} within the algorithm,
 and the operations done here are specified by the algorithm's input (the function $f$). 
The general treatment of post-quantum oracles is subtle~\cite{LeeB15}.
For instance, in the circuit formalism of the Deutsch--Jozsa algorithm, 
 one uses a control gate incorporating {\em phase kickback}---%
 a phenomenon in quantum theory where the transformation induces a locally-unobservable phase on the system it acts on, but this phase takes on operational consequence when considered in the context of a coherently controlled operation.
Requiring the existence of such phase kickback in general restricts the theories that may be used~\cite{LeeS16}.
However, the interferometric set-up considered in this article is a single (not multi-partite) system, 
 and all relevant phases are operationally represented by this system's state.
Thus, we do not require the full formalism of post-quantum oracles here, 
 but instead focus on the properties a theory requires to perform computation specifically using a branching interferometer.

We wish to identify theories that allow us to find interferometric oracles for {\em all} possible functions $f:\mathcal{B}^n\to \mathcal{B}$,
 and then from the output of this oracle, determine the answer to the constant vs.\ balanced problem in a systematic manner.
In particular, we shall consider a distributed regime where at each of the $2^n$ branches of the interferometer there is an independent agent,
 and each locally calculates $f(x)$ and chooses a phase plate to add to their branch based on the result of their calculation.
(Equivalently, we could think of the $2^n$ branches of the interferometer as encoding a lookup table for the outputs associated with each input to $f$.)
The transformation induced by agent $x$'s phase-plate should not depend on the outcomes of $f(y\neq x)$ obtained by the other agents,
 but it might be dependent on which branch the agent is in.
Crucially, we require that the transformation induced by each agent is compatible with {\em branch locality} on that branch.

For a theory to implement such an oracle query as an interferometer, we suggest the following criterion:
\begin{enumerate}[i.]
\item For every bit-string $x\in \mathcal{B}^n$, there must be at least two choices of transformations $T^1_x$ and $T^0_x$ that can be applied locally to branch $x$, which may be associated with $f(x)=1$ and $f(x)=0$ respectively.
\end{enumerate}
If the theory does not admit both 
 $T^1_{\tilde{x}}$ and $T^0_{\tilde{x}}$ for some value of $\tilde{x}$, 
 then there could be no operational distinction between the oracles corresponding to functions $f$ and $g$ where $f(\tilde{x})=0$ and $g(\tilde{x})=1$ but $f(x)=g(x)$ for all other $x\neq \tilde{x}$.

The input to the algorithm (the function $f$) is mapped onto a particular choice of oracle.
We require that the other aspects of the circuit (namely, the state input to the oracle; and the choice of measurement that extracts statistics from the oracle's output state) are hence independent of the algorithm's input $f$.
In quantum theory, this would amount to using the same set of beamsplitters for every choice of $f$.
This is expressed by our second criterion:
\begin{enumerate}[i.]
\setcounter{enumi}{1}
\item There must be some effect $\covec{e}_\mathrm{C}$ such that when the output state $\vec{s}_\mathrm{out}$ is measured, $\covec{e}_\mathrm{C}\cdot\vec{s}_\mathrm{out} = 1$ if $f$ is constant, and $\covec{e}_\mathrm{C}\cdot\vec{s}_\mathrm{out} = 0$ if $f$ is balanced, for at least one input state $\vec{s}_\mathrm{in}$ chosen independently of $f$.
\end{enumerate}

The statement of this criterion is specific to the constant vs.\ balanced problem. 
Related criteria could be formulated for other computational tasks (such as Grover's algorithm, as we shall discuss in \cref{sec:Grover}), whereby the conditions on $\covec{e}_\mathrm{C}$ are slightly different.
However, the criterion for a different problem solvable by an interferometric computation should maintain independence from the computation's input $f$,  and the choices of state input to the oracle and the measurement made on that oracle's output.

Note that here we have strictly required a deterministic output to the measurement (in direct analogy with the quantum Deutsch-Jozsa protocol): there must be a single measurement that determines whether $f$ is constant or balanced. 
A slightly looser requirement would be that there is some $\covec{e}_\mathrm{C}$ such that $\covec{e}_\mathrm{C}\cdot\vec{s}_\mathrm{out} > \frac{1}{2}$ if $f$ is constant, and $< \frac{1}{2}$ if $f$ is balanced, such that the two classes of $f$ may ultimately be distinguished by enough repetitions of the experiment.

% QUANTUM THEORY
\inlineheading{Quantum theory.}
As expected, quantum theory is compatible with both requirements.
In general, an agent $x$ can implement $U^1_{x} = -\ketbra{x}{x} + \sum_{i\neq x} \ketbra{i}{i}$ if $f(x)=1$ and $U^0_{x} = \sum_{i\in\mathcal{B}^n} \ketbra{i}{i} = \id$ otherwise.
Transformations of this form are compatible with branch locality (proven explicitly in \cref{app:BLQT}), and can be composed in any order (the operations are all diagonal in the same basis) such that the overall effect of all the agents is to induce the operation $U_f = \sum_{x\in\mathcal{B}^n} e^{i\pi f\left(x\right)} \ketbra{x}{x}$.
By preparing the initial state $\ket{+}^{\otimes n}$, one can perform a final measurement in the $X\otimes\cdots\otimes X$ basis (i.e.\ by applying $H^{\otimes n}$ followed projecting into $\ket{0\ldots0}$).
This has probability of a positive outcome given by $p = \frac{1}{2^n} |\sum_{x\in\mathbb{B}^n} e^{i\pi f\left(x\right)} |^2$.
As $f(x)=0$ contributes a term $e^0=1$ and $f(x)=1$ contributes $e^{i\pi}\!=\!-1$,
we see that $p\!=\!1$ when $f$ is constant as the global sign does not matter, and $p\!=\!0$ for any balanced $f$ because half the addends will cancel out the other half.

% CLASSICAL THEORY
\inlineheading{Classical theory.}
On the other hand, classical theory does not permit this algorithm, but rather fails at requirement i since it does not admit any non-trivial (i.e.\ other than $\id$) phase transformations~\cite{GarnerDNMV13}, and so $T_\mathrm{x}=\id$ always.
There will therefore be no way to encode any measurable information about $f$ through actions taken on the separate branches, and so the final measurement will only depend on the initial state $\vec{s}_\mathrm{in}$.

% BOX-WORLD
\inlineheading{Box-world single systems (gbits).}
We can confirm the suggestion made in \cite{DahlstenGV14} that if the ``which branch?'' freedom of an interferometer is described by a gbit 
 (where one of the gbit's principle measurements corresponds to the traversing particle's position), 
 then this system cannot implement any interferometric computation\footnote{
It should also be noted that implementing the Deutsch--Jozsa algorithm to solve this problem on a collection of gbits (i.e.\ in the circuit picture) can also not be done in box-world, because the $f$-controlled-NOT gate will not be an allowed transformation, according to the restrictions on correlating dynamics shown in~\cite{GrossMCD10}.
}.
In particular, criterion i fails because {\em no} transformation can be localized to {\em any} subset of branches (proven for the general case in~\cite{DahlstenGV14}) without violating branch locality in some way, and so the agents on the separate branches have no way of encoding the outcome of $f(x)$. 

Let us see this explicitly for the $n=1$ case, where the ``which-branch?''\ measurement is a binary degree of freedom and the family of gbits are $d$-dimensional hypercubes. 
As $d=1$ is just classical theory, we shall consider the simplest non-trivial example, $d=2$. 
The state-space of this theory is given by a square (sometimes called a {\em square bit}), and the phase-group is $Z_2$, consisting of either doing nothing ($\id$), or exchanging the statistics of the complementary measurement ($X$-flip).
Writing the position and complementary measurements as $Z$ and $X$, we note that any operation branch local to $Z=1$ must not change states where $P(Z=1)=0$.
However, as the action of $X$-flip is independent of $Z$, there is a state where $Z=-1$ and $X=1$ which will be turned into one where $Z=-1$, $X=-1$ when the $X$-flip is applied.
Thus, the $X$-flip operation cannot be applied branch locally at $Z=1$. 
Similar logic rules it out localization to $Z=-1$, and hence we see that although $X$-flip is in the global phase group, it cannot be assigned to either branch.
The only remaining group element is $\id$, and this does nothing, and hence the localizable phase dynamics are trivial.

If it were possible to apply $\id$ and $X$-flip from each branch, then the algorithm could be performed as an interferometric computation.
However, as they cannot, such a calculation can only be done by global phase, in which $\id$ is applied globally if the function is constant, and $X$-flip if it is balanced (and then $\covec{e_C} = \covec{x}_+$ such that $\covec{x}_+\cdot\vec{s} = P(X\!=\!+1)$ could satisfy criterion ii).
This effectively reduces the problem to a trivial oracle that answers the question ``is the function constant or balanced?''\ and no longer has the structure that allows the problem to be treated as a distributed computation.

That there is a global implementation, but not a local one, highlights a key difference between interference in box-world and interference in quantum theory.
Processes that induce phase shifts in an intrinsically non-branch-local way (such as by the Aharonov--Bohm effect~\cite{AharonovB59} in which switching on a magnetic solenoid between two branches induces a geometric phase) may still be plausible in box-world.
In quantum theory, any phase operation induced by the Aharonov--Bohm effect could be equivalently induced by pieces of glass placed in each branch: there are no operations in the phase group $\mathcal{G}$ (in this case $\bigoplus^N\!\Un{1}$) that are not in the union of all the operations local to each branch $\bigcup_i \mathcal{G}_i$.
On the other hand, in box-world $\bigcup_i \mathcal{G}_i = \{\id\}$, and so even if there is a process that can induce a non-trivial global phase (such that $\mathcal{G}\neq\{\id\}$) there is no way to generate this phase from local operations.
Thus, whilst quantum theory could do the algorithm distributing $f$ over many branches, the best that is admissible in box-world would constitute of a single agent who evaluates all values of $f$ to see if it is constant or balanced, and then applies the global operation (e.g.\ switches on a solenoid, or some analogue thereof) only if it is constant.

% N-BALLS
\inlineheading{$d$-balls and quaternionic quantum theory.}
A $d$-ball is the generalization of the single-qubit Bloch-sphere into $d$ dimensions, by generalizing the three binary measurements into $d$ such measurements~(see e.g.\ \cite{DakicB13,MullerM13,BarnumMU14,GarnerMD14}).
For states with $d$ measurements $\{X_i\}$, the state-space satisfies a quantum-like uncertainty relation given by $\sum_i \left(P(X_i = 1) - \frac{1}{2}\right)^2 \leq \frac{1}{4}$, such that if any measurement has a certain outcome, all others are uniformly random.
Geometrically, this corresponds to a $d$-dimensional hypersphere. 
Excluding reflections (by analogy with quantum theory), the automorphism group of a single $d$-ball state-space is given by $\SO{d}$, and the phase group after fixing the values of one measurement is $\SO{d-1}$.

Let us first consider single-bit systems.
First, we note that {\em all} operations in the phase group can be implemented locally on either branch.
Intuitively, this is because for each branch, there is a unique state that has no support on that branch, and as this state has uniformly random outcomes for every measurement except the position one, it is invariant under application of $\SO{d-1}$, and thus will never violate the requirement of branch locality.
Ignoring $d=1$ as it is classical, and $d=2$ (because it only has a non-trivial phase group if reflections are permitted), we see that for $d>2$, we can pick a $\pi$ (generalized) rotation
about the $Z$ axis and each agent can apply the same rotation when $f(x)=1$ (and do nothing when $f(x)=0$).
The algorithm will then work exactly as the $1$-bit quantum case did; and there will be a measurement whose outcome changes from one value to another if $f(0)\neq f(1)$.
(As we only need the identity and one other operation, which forms an Abelian subgroup isomorphic to integer addition modulo 2 ($Z_2$), the issues of commutativity addressed in \cite{GarnerMD14} should not cause problems in the implementation of this algorithm.) 

The general algorithm is trickier to discuss, as the spherical shape of the state-space is specific to a two-level system (and thus generalizations of this will not provide obvious insight as to what measurements and transformations are possible on a $N=2^n$ level system).
However, the $d$-balls are in general the state-spaces associated with two-level systems described by a Jordan algebra~\cite{AlfsenS03}, which includes real-valued ($d=2$), complex ($d=3$) and quaternionic~\cite{Graydon11} ($d=5$) quantum theory (i.e.\ variants of quantum theory where the Hilbert space is on a different field).
Thus, at least in these cases, we might attempt to address the $n$-bit problem using the appropriate algebra.
(By using the many-level single-system approach, we also dodge the bullet of attempting to compose a multipartite system of $n$ $d$-balls, where in general it is tricky to find the joint state-space since the tensor-product of such state-spaces is not uniquely defined.)

Let us then single out the example of quaternionic\footnote{
 Quaternions may be thought of as the extension of the field of complex numbers~\cite{Dickson19} to include three different imaginary elements $i$, $j$ and $k$ such that $ii = jj = kk = ijk = -1$. See~\cref{app:Quart} and \cite{Graydon11,Baez14}.} quantum theory.
Beyond the operational probability vector representation appropriate for any GPT  (e.g.\ the two-level quaternionic state-space is the $5$-ball),
 states in quaternionic quantum theory may also be represented similarly to those in complex quantum theory:
 as both positive semi-definite density matrices in general,
 or as quaternionic vectors $\mathbb{H}^N$ for pure states.
The reversible operations on a quaternionic state are elements of the symplectic group $\Sp{N}$~\cite{Baez14}.
We may express pure quaternionic states using the bra-ket notation, noting that the inner product between bras and kets is now the symplectic product (i.e.\ to turn a quaternionic ket into a bra, we must transpose and apply {\em quaternionic conjugation} to it~\cite{Graydon11}).

As shown in shown \cref{app:Quart}, an $N$-level quaternionic system contains $\bigoplus^N \Sp{1}$ in its phase group (where $\Sp{1}$ are the unit quaternions; isomorphic to $\SU{2}$).
Moreover each $\Sp{1}$ addend corresponds to operations that can be applied locally on a particular branch, and quaternionic states that different only by a sign are operationally indistinguishable (i.e.\ there is a $Z_2$ {\em global phase}, much like the $\Un{1}$ complex quantum global phase).

In the context of evaluating $f$, each agent could thus choose between implementing $\id$ or $-1 \in\Sp{1}$ (where $-1 \cdot -1 = \id$ and corresponds to changing the quaternionic phase of a possibility by a real factor of $-1$) on their branch depending on the outcome of $f$.
One can prepare the quaternionic superposition state $\frac{1}{\sqrt{N}} \sum_j \ket{j}$ (much like in complex quantum theory), and the action of the agents over the $N$ branches will be to introduce phase of $-1$ in front of either all, none or exactly half of these terms.
Because of the $Z_2$ global phase, the two balanced functions will result in operationally indistinguishable states; and in general, any function $f$ could be replaced by its negation without changing the statistics output by the interferometer\footnote{
In general balanced functions which are not negations of each other result should result in operationally distinguishable statistics for some measurements, even if they all satisfy $\covec{e_C}\cdot\vec{s} = 0$.
This can be conceptually understood by discarding half of the branches and performing interferometry on the remaining half: the choice of branches included in the subset in will general govern whether the subset is balanced, constant or neither, and if it was one of the first two, then a lower-dimension version of the algorithm should work on these.
It would thus be disingenuous if all balanced functions resulted in the same statistics.
}.

Finally, we must argue for the existence of a quaternionic beamsplitter whose effect is to sum together all these terms.
This can be seen by noting that $\Sp{N}$ is isomorphic to $\Un{N, \quat}$, which contains the usual quantum Hadamard gate that only contains real numbers and is sufficient for this task.
(In general, there will be more types of quaternionic beamsplitters than of quantum beamsplitters; but we only need one for this algorithm).
Thus requirement ii is also satisfied, and we hence conclude that an interferometric computation to solve the constant/balanced problem can be done in quaternionic quantum theory.

\begin{figure}[tbh]
\begin{center}
\includegraphics[width=0.85\textwidth]{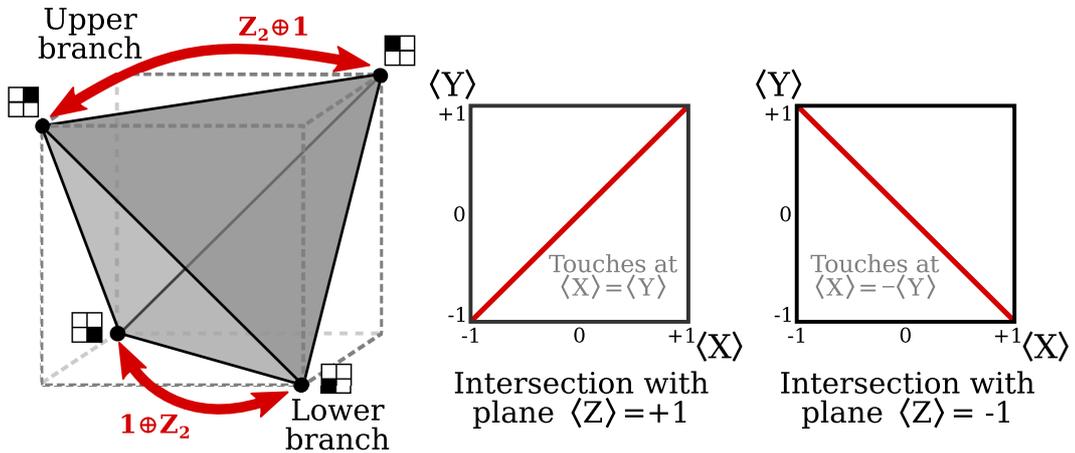}
\caption[Localisable dynamics for the ontic state-space of Spekkens' toy theory]{
\textbf{Localisable dynamics for the underlying hidden variable state-space of Spekkens' toy theory.}
The phase group $Z_2\oplus Z_2$ splits into a different subgroup for each branch (drawn on LHS).
For the lower branch, there is a line of states with support on this branch: the states on the upper branch, drawn in the middle figure. Branch locality mandates that each of these states must not be altered by a transformation on the lower branch, and so one can either do nothing or exchange $\spek{3}$ and $\spek{4}$, resulting in the group $\id\oplus Z_2$.
Similarly, for the upper branch, states with no support (RHS) must not be changed, and this allows only for doing nothing or 
exchanging $\spek{1}$ and $\spek{2}$, yielding the group $Z_2 \oplus \id$.
}
\label{fig:SpekOntic}
\end{center}
\end{figure}

% SPEKKENS' TOY MODEL
\inlineheading{Spekkens' toy model.}
Let us consider the $1$-bit implementation of the problem in the context of Spekkens' toy model.
By comparison of the behavior of Spekkens' toy model and the underlying hidden variable without the epistemic restriction, we can see explicitly how the uncertainty principle makes the algorithm possible.
The phase group reversible operations that preserve the $Z$ measurement of Spekkens' toy model is $Z_2\oplus Z_2 = \{ 1234, 2134, 1243, 2143\}$ (using the one-line permutation notation).
By application of branch locality to the state-space of the underlying hidden variable, we find that the upper branch can locally execute the operations $Z_2\oplus\id = \{1234, 2134\}$ and the lower branch can locally execute $\id\oplus Z_2 = \{1234,1243\}$ (and if both branches implement a non-trivial transformation the final element $2143$ is performed).
This may be seen visually in \cref{fig:SpekOntic}.

%% FIGURE: Spekkens state-space

Without the epistemic restriction,
 the operations on the two branches correspond to two disjoint permutations.
Thus, there is no way for any action done by an agent on one branch to be cancelled out by an action done by the agent on the other.
However, the single-bit Deutsch-Jozsa algorithmic is effectively a distributed modulo-$2$ addition.
There will be no way for constant and balanced $f$ to result in mutually exclusive output states.
For example, suppose $\spek{13}$ was input, and each agent applies $\id$ if $f(x)=0$, and the non-trivial element otherwise; 
the outcomes when the function is constant will be $\spek{13}$ or $\spek{24}$, and if the function is balanced will be $\spek{14}$ or $\spek{23}$.
This failure is more subtle than that of box-world: here, we can find an encoding of $f(x)$ outcomes to operations on branch $x$ that seemingly satisfies requirement i, but there is no such choice of encoding that allows an appropriate effect to be found that satisfies requirement ii.
No single measurement can distinguish between these two cases, and one would have to perform complete tomography on the state (hence identifying $f$) before one could say if $f$ were constant or balanced.
This fails even the weaker version of requirement ii: no single effect has $\covec{e}_\mathrm{C}\cdot\vec{s} > \frac{1}{2}$ if and only if $f$ is constant.
One might interpret this as there being a complete lack of global phase in the ontic state-space, such that the action on one branch cannot cancel out with the action taken on another (cf.\ complex and quaternionic quantum theory where $\diag{\left(-1,1\right)}$ and $\diag{\left(1,-1\right)}$ are equivalent to each other, since they only differ by a global phase).

This is no longer the case when we apply the epistemic restriction 
 that turns the classical variable into Spekkens' toy model.
The state-space is now subject to a quantum-like uncertainty principle such that if the outcome of $Z$ is known with certainty, no degrees of freedom remain.
This implies that branch locality no longer divides the phase group into two separate subgroups, but rather admits for the application of any element in $Z_2\oplus Z_2$ from either branch.
Thus both agents could choose (for example) to execute $1234$ when $f(x)=0$ and $2143$ when $f(x)=1$.
These operations will combine to form a representation of $Z_2$, and hence the single-bit Deutsch algorithm will be possible once more.
For example, $\spek{13} \to \spek{13}$ when $f$ is constant and $\spek{13}\to\spek{24}$ when $f$ is balanced, which can be distinguished through a measurement of $X$.

In terms of the hidden variable, allowing for these operations leads to remarkable behaviour.
If one were to interpret $\spek{1}$ and $\spek{2}$ as lying definitely with one agent and $\spek{3}$ and $\spek{4}$ with the other (reasonable under branch locality; see \cref{fig:SpekOntic}) then with the epistemic restriction, an agent can seemingly violate branch locality on the hidden variable, and change the state of the system even when it has no support in her branch.
However, because of the epistemic restriction, this violation can never be observed.
A similar phenomenon can be observed in another famous hidden variable theory: Bohmian mechanics~\cite{Bohm52,Bohm52b,Norsen14}.
In Bohmian mechanics it is possible~\cite{Norsen14} to construct a set-up involving an entangled pair of particles entering Stern--Gerlach devices at two different positions in space, such that if the exact initial position of the one of the particles was known, at one detector it would be possible to determine from the final trajectory taken by the particle whether or not a measurement had made by the remote detector, violating no-signalling.
However, because knowing the initial position is forbidden, this violation is avoided by an epistemic restriction.

%%
%% GROVERS' ALGORITHM
\section{Unordered database search: Grover's algorithm}
\label{sec:Grover}
A classical search to find one record in an unordered database containing $N$ entries requires on average $N/2$ queries, whereas Grover's algorithm~\cite{Grover96} provides a quantum method which can determine this after $\mathcal{O}\left(\sqrt{N}\right)$ queries.
Much like the Deutsch--Jozsa algorithm, one can implement the Grover algorithm without recourse to entanglement by building a sufficiently large single-system set-up~\cite{Lloyd99}.
In such a set-up, one passes a single particle through a series of beamsplitters such that it is in a superposition over all $N$ branches.
These $N$ branches are connected to the database, which could be thought of as $N$ black boxes, one containing a phase plate that induces a $\pi$ phase shift, and the rest do not change the phase.
The database could be thought of as a function $f(x)=\{0,1\}$ where $f(x)=1$ for only one value of $x$, and $f(x)=0$ otherwise.
Following this, the output branches are passed through a second set of beamsplitters, and an additional phase-shift of $\pi$ is added to the first branch only.
This entire process is repeated $\sqrt{N}$ times, after which it is highly likely  ($p>\frac{1}{2}$) that the particle will be in the branch corresponding to where the $\pi$ phase plate is.

Much like the many-branch implementation of constant vs.\ balanced problem, a larger number of optical components are required to prepare the superposition states of an $N$ level system.
However, if the resource to be minimized is the number of times the database can be probed (e.g.\ because it is far away; or charges per access, 
 whereas interferometric apparatus is free), then the $N$-branch implementation (as opposed to the qubit implementation) of this algorithm is just as efficient. 

The unordered search addressed by Grover's algorithm is slightly less straightforward to classify than the constant vs.\ balanced problem.
Certainly, we will still need requirement $i$ to be true in order to build our database; in this case the function $f$ is no longer constant or balanced, but rather guaranteed to only be $1$ for one value of $x$.
Thus, every theory which failed on this requirement (classical bits, boxworld) will also be unable to provide the coherent oracle required for Grover's algorithm.
In the box-world case, as with the constant vs.\ balanced problem, one can only admit a ``global oracle'' which trivially sets the state such that it flags the correct value of $x$ satisfying $f(x)=1$).
On the other hand, quaternionic quantum theory admits all the components required (local phase-shifts and beamsplitters), which will work in the same way as in quantum theory, and so should allow an implementation of Grover's algorithm.

In this article, we considered the post-quantum implementation of Grover's algorithm explicitly as a many-branch interferometer, with the physical constraints that this entails.
Other works have focused on the {\em computational efficiency} of similar algorithms in post-quantum theories.
\citet{FernandezS03} demonstrate that any complex quantum circuit can also be efficiently implemented using ``real bits'' (qubits disallowing complex numbers-- that is, the Jordan algebra with circular state-spaces) efficiently; and that quaternionic circuits may likewise be efficiently simulated by quantum bits (provided one picks an ordering for the quaternionic gates to be simulated).
This would imply that Grover's algorithm is as good in these theories as quantum theory, provided that the circuit can be realised and an appropriate oracle defined (see discussion in~\cite{LeeB15}).
Grover's algorithm has also been considered in the context of Sorkin's hierarchy of interference by~\citet{LeeS16b}.
The authors show, in the context of additional assumptions, that allowing for higher order interference effects than possible in quantum theory (e.g.\ theories where three-slit interference patterns can't be explained as mixtures of pair-wise interferences, unlike quantum theory~\cite{Sorkin94}) does not improve on the  $\mathcal{O}\small(\sqrt{N}\small)$ efficiency possible in quantum theory.

%%
%% SECTION: Discussion
\section{Summary}
By treating spatial degrees of freedom in the framework of generalized probabilistic theories, we have seen that interferometric computation can be extended beyond quantum theory.
The usefulness of a theory was shown to be very much reliant on the structure of available dynamics.
In particular, locality considerations dictate whether the process can be interpreted as a distributed computation performed by many agents.
We ruled out box-world as the dynamics could not be localized to different branches, and so the only interferometric solution to either the constant vs.\ balanced or the unordered search problems would require a trivial oracle that directly answers the question.

On the other hand, in theories with interesting dynamics interferometric computation is once more possible.
In complex and quaternionic quantum theory, which both contain a non-trivial unmeasurable global phase group that is also contained within every branch local subgroup, we saw that it is possible for the agents to act on separate branches in such a way that both of the constant functions produce the same operational output state.

Finally, when we considered Spekkens' toy hidden variable model, we found that the computation was made possible explicitly by the imposition of an uncertainty principle, which censored the measurement of non-local behavior in the underlying underlying state, in a manner that is similar to the censorship of locality violations arising from hidden variables in Bohmian mechanics.

%% ACKNOWLEDGEMENTS
\section*{ACKNOWLEDGEMENTS}
The author is very grateful for illuminating discussions and correspondence with \mbox{Felix} \mbox{Binder}, Oscar Dahlsten, Daniela~Frauchiger, Nana Liu, Markus M\"uller, \mbox{Vlatko Vedral}, and \mbox{Benjamin} Yadin.
The author is grateful for financial support from the John Templeton Foundation and the Foundational Questions Institute.
This research was undertaken whilst the author was funded by the Engineering and Physical Sciences Research Council (UK) at the University of Oxford.
Some of concepts in this article have also appeared as part of the author's DPhil thesis~\cite{Garner15DPhil}.

%% BIBLIOGRAPHY

\newpage
\appendix

\section{The uncertainty relation, and the qubit state-space}
\label{app:Uncertainty}

The {\em Cauchy--Schwarz inequality}, as applied to complex Hilbert spaces, states that 
\begin{equation}
\label{eq:CSineq}
\braket{\xi}{\xi}\braket{\chi}{\chi} \geq |\braket{\xi}{\chi}|^2
\end{equation}
for complex vectors $\ket{\xi}$ and $\ket{\chi}$ (see, e.g.\ \cite{NielsenC00}).
From this, we can derive a bound on the product of variances of measurement operators---that is, the Schr\"odinger--Robertson uncertainty relation~\cite{Schrodinger30,Robertson29}---and the spherical shape of the qubit state-space when expressed as the expectation values of Pauli operators.

Consider a pair of operators $A$ and $B$ both acting on some quantum state $\ket{\psi}$.
Ineq.~\eqref{eq:CSineq} implies that
$|\bra{\psi} A B \ket{\psi}|^2 \leq \bra{\psi} A\herm A\ket{\psi} |\bra{\psi} B\herm B \ket{\psi}$ for any $\ket{\psi}$.
When $A$ and $B$ are also Hermitian, then
\begin{equation}
\label{eq:CSherm}
|\bra{\psi} A B \ket{\psi}|^2 \leq \bra{\psi} A^2\ket{\psi} \bra{\psi} B^2 \ket{\psi}.
\end{equation}
For Hermitian $A$ and $B$, $\left(\bra{\psi}AB\ket{\psi}\right) \herm = \bra{\psi}BA\ket{\psi} = \bra{\psi}AB\ket{\psi}^*$, and so $\bra{\psi}(AB+BA)\ket{\psi}$ is a real number equal to twice the real part of $\bra{\psi}AB\ket{\psi}$.
Similarly, $\bra{\psi}(AB-BA)\ket{\psi}$ is purely imaginary, equal to twice the imaginary part of $\bra{\psi}AB\ket{\psi}$. 
Using $|\bullet|^2 = \Real\left({\bullet}\right)^2 + \Img\left({\bullet}\right)^2$, we thus find that 
$4|\bra{\psi}AB\ket{\psi}|^2 = |\bra{\psi}(AB+BA)\ket{\psi}|^2 + |\bra{\psi}(AB-BA)\ket{\psi}|^2$.
Substituting this in ineq.~\eqref{eq:CSherm}:
\begin{equation}
\label{eq:commuteUnc}
|\bra{\psi} [A, B] \ket{\psi}|^2 
+ |\bra{\psi} \{A, B\} \ket{\psi}|^2 
\leq 
4 \bra{\psi} A^2\ket{\psi} |\bra{\psi} B^2 \ket{\psi},
\end{equation}
where we have written the {\em commutator} $[A,B] = AB - BA$ and the {\em anti-commutator} $\{A, B\} = AB + BA$.

Suppose we pick some state $\ket{\psi}$, 
 and two Hermitian observables $X$ and $Y$, 
 then the matrices $A =: X - \expt{X} \id $ and $B =: Y-\expt{Y} \id$ (where $\expt{\bullet} := \bra{\psi}\bullet\ket{\psi}$ is the expectation value of measurement $\bullet$ of state $\ket{\psi}$) are also Hermitian.
We may evaluate
\begin{align}
[A, B] & = [X, Y], \\
\{A, B\} & = \{X, Y\} -2\left(X\expt{Y}+Y\expt{X}\right) + 2\expt{X}\expt{Y}\id,  \\
A^2 & = X^2 - 2X\expt{X} + \expt{X}^2\id, \\
B^2 & = Y^2 - 2Y\expt{Y} + \expt{Y}^2\id.
\end{align}
Taking the expectation value of the above on the same state $\ket{\psi}$ yields:
\begin{align}
\expt{[A,B]} & = \expt{[X, Y]}, \\
\expt{\{A,B\}} & = \expt{\{X,Y\}} - 2\expt{X}\expt{Y}, \\
\expt{A^2} & = \expt{X^2} - \expt{X}^2, \\
\expt{B^2} & = \expt{Y^2} - \expt{Y}^2.
\end{align}
The last two expressions are the variances $\Delta M^2 : = \expt{M^2} - \expt{M}^2$ of $X$ and $Y$ respectively.

Substituting these into the Cauchy--Schwarz inequality (ineq.~\eqref{eq:commuteUnc}) leads us then to Schr\"odinger's statement of the uncertainty relation~\cite{Schrodinger30}:
\begin{eqnarray}
\label{eq:UncertSchro}
\frac{1}{4} |\expt{[X, Y] }|^2  + \frac{1}{4} |\expt{\{X,Y\}} - 2\expt{X}\expt{Y}|^2 & \leq & \Delta X ^2 \Delta Y^2.
\end{eqnarray}

This also implies the looser bound of the Robertson uncertainty principle~\cite{Robertson29}, because
 $|\bra{\psi}\{A,B\}\ket{\psi}|^2$ 
 in ineq.~\eqref{eq:commuteUnc} is never negative:
\begin{eqnarray}
\frac{1}{4}|\expt{[X, Y] }|^2 & \leq & \Delta X ^2 \Delta Y^2.
\end{eqnarray}

We may specialize the Schr\"odinger uncertainty principle of ineq.~\eqref{eq:UncertSchro} to a two-level quantum system (qubit), 
 and choose two of the Pauli matrices
\begin{equation}
\left\{
\sigma_x = \left(\begin{array}{cc}
0 & 1 \\
1 & 0 \end{array}
\right), \quad
\sigma_y = \left(\begin{array}{cc}
0 & -i \\
i & 0 \end{array}
\right), \quad
\sigma_z = \left(\begin{array}{cc}
1 & 0 \\
0 & -1 \end{array}
\right)
\right\}
\end{equation}
(say $\sigma_x$ and $\sigma_y$) as $X$ and $Y$.
Using $[\sigma_x, \sigma_y] = 2i\sigma_z$, $\{\sigma_x,\sigma_y\}=0$, $\Delta X^2 = 1-\expt{\sigma_x}^2$, and $\Delta Y^2 = 1-\expt{\sigma_y}^2$, we find
\begin{equation}
 \expt{\sigma_z}^2+\expt{\sigma_x}^2\expt{\sigma_y}^2 \leq (1-\expt{\sigma_x}^2)(1-\expt{\sigma_y}^2),
\end{equation}
which rearranges to
\begin{equation}
\expt{\sigma_x}^2 + \expt{\sigma_y}^2 +\expt{\sigma_z}^2 \leq 1.
\end{equation}
Thus, when the qubit state-space is paramaterized in terms of the expectation values of Pauli matrices,
 the expectation values of pure states are bound by a sphere of unit radius.
Closing these pure states into a convex set (i.e.\ by allowing mixtures), we find that {\em all} qubit states must be within the {\em Bloch sphere}.

\vspace{1em}

\section{Phase and branch locality in quantum theory}
\label{app:BLQT}
First, we show that the phase group of a projective measurement on an $N$-level quantum system is given by $\bigoplus^N \Un{1}$.

\begin{lemma}
\label{lem:RevQuantPhase}
For an $N$-level quantum system, with a projective measurement $Z$, whose outcomes may be expressed in terms of projectors $\big\{Z_j =\ketbra{j}{j}\big\}_{j=1\ldots N}$ (and $\sum_j Z_j = \id$),
all reversible phase operations (i.e.\ satisfying \cref{def:phase}) may be written in the form
\begin{equation}
U = \sum_j e^{i \phi_j} \ketbra{j}{j}.
\end{equation}
Conversely, all matrices of this form are a phase operation with respect to $Z$. 
\begin{proof}
We will use the isomorphism between states $\vec{s}$ and density matrices $\rho$, and between effects $\covec{e}$ and measurement operators $M$, such that the inner product in both cases gives the associated probability of a particular outcome ($\covec{e}\cdot\vec{s} \leftrightarrow \tr\left(M\rho\right)$), and the fact that general reversible quantum transformations act as $\rho \to U \rho\, U\herm$ for $U\in\Un{N}$.
This allows us rewrite the condition for a reversible phase operation with respect to $\{\covec{z_i}\}$ (written in the GPT framework as $\covec{z_j}\cdot T \vec{s} = \covec{z_i}\cdot\vec{s} \quad \forall j, \vec{s}$) as an equivalent statement in the quantum formalism:
\begin{equation}
\label{eq:QPhaseTransForm}
\tr\left(Z_j  U\rho\,U\herm \right) = \tr\left(Z_j \rho \right) \qquad \forall \rho, j.
\end{equation}
This implies that $[U, Z_j] = 0$ for all $j$, and so we can write $U$ diagonally in the basis of $Z$,
\begin{equation}
\label{eq:qphasetrans}
U = \sum_j e^{i \phi_j} \ketbra{j}{j}.
\end{equation}

To show the converse, because $U$ of the above form commutes with all elements $Z_j$, and $U\herm U =\id$,  we may use the cyclicity of trace to write:
$\tr\left(Z_j U \rho\, U\herm \right) = \tr\left(U\herm Z_j \, U\rho \right) = \tr\left(U\herm U Z_j\rho \right) =\tr\left(Z_j \rho \right)$, which is exactly the phase condition written in \cref{eq:QPhaseTransForm}.
\end{proof}
\end{lemma}

From this, we can show the form an operation local to one branch must take agrees with~\cref{eq:quantumBL}:

\begin{lemma}
Branch locality implies that the change induced by a reversible quantum operation $U_i$ on a branch $i$ in a set of disjoint branches can be most generally expressed by a complex phase acting only on the ket $\ket{i}$ associated with being in that particular branch:
\begin{equation}
\label{eq:BLU}
U_i = e^{i\Phi} \Big( e^{i\phi_i} \ketbra{i}{i} + \sum_{j\neq i} \ketbra{j}{j} \Big).
\end{equation}

\begin{proof}
For a branch $i$, and a state $\rho_{\lnot i}$ that has no support on $i$ such that
\begin{equation}
\label{eq:cond:nosupport}
\tr\left(\rho_{\lnot i} Z_i\right) = 0,
\end{equation}
then a branch local operation $U_i$ on branch $i$ always satisfies
\begin{equation}
\tr\left(U_i \rho_{\lnot i} U_i\herm M \right) = \tr\left(\rho_{\lnot i} M \right)
\end{equation}
for any observable $M$.
This implies that 
\begin{equation}
\label{eq:cond:BLQM}
U_i \rho_{\lnot i} U_i\herm = \rho_{\lnot i}.
\end{equation}

The most general state that satisfies \cref{eq:cond:nosupport} may be written in the $Z$ basis as
\begin{equation}
\label{eq:nosuppstate}
\rho_{\lnot i} = \sum_{\alpha,\beta\neq i} \rho_{\alpha\beta} \ketbra{\alpha}{\beta}.
\end{equation}

According to \cref{lem:RevQuantPhase}, we can express a general quantum phase transformation with respect to $Z$ as 
$U = \sum_j e^{i \phi_j} \ketbra{j}{j}$,
and consider its action on $\rho_{\lnot i}$:
\begin{eqnarray}
U_i \rho_{\lnot i} U_i\herm & = & \sum_j \sum_{\alpha,\beta\neq i} \sum_k e^{i \phi_j} \rho_{\alpha\beta} e^{-i \phi k} \ketbra{j}{j}\chainketbra{\alpha}{\beta}\chainketbra{k}{k} \\
& = & \sum_{\alpha,\beta\neq i} e^{i \left( \phi_\alpha - \phi_\beta\right)} \rho_{\alpha\beta} \ketbra{\alpha}{\beta}.
\end{eqnarray}
Therefore, for the condition in \cref{eq:cond:BLQM} to be satisfied, we require
\begin{eqnarray}
\label{eq:rev_bl_qm_cond}
\sum_{\alpha,\beta\neq i} e^{i \left( \phi_\alpha - \phi_\beta\right)} \rho_{\alpha\beta} \ketbra{\alpha}{\beta} = \sum_{\alpha,\beta\neq i}  \rho_{\alpha\beta} \ketbra{\alpha}{\beta}.
\end{eqnarray}
For this to hold, it must be true for every element $\ketbra{\alpha}{\beta}$, and this requires $\phi_\alpha = \phi_\beta$ (or for $\phi_\alpha$  and $\phi_\beta$ to differ by an unmeasurable multiple of $2\pi$) for all $\alpha \neq i,\beta \neq i$.
A transformation $U_i$ local to branch $i$ thus has the form
\begin{equation}
\label{eq:BLU}
U_i = e^{i\Phi} \Big( e^{i\phi_i} \ketbra{i}{i} + \sum_{j\neq i} \ketbra{j}{j} \Big)
\end{equation}
where $\Phi$ is some global phase that can be ignored.

\end{proof}
\end{lemma}

\section{Phase and branch locality in quaternionic quantum theory}
\label{app:Quart}
By using insights from Graydon~\cite{Graydon11}, we can generalise the results of \cref{app:BLQT} into quaternionic quantum theory.
A quaternion $h \in \quat$ may be written $h = a + ib + jc + kd$ where $a,b,c,d \in \mathcal{R}$ and $i$, $j$ and $k$ are three types of imaginary number satisfying $ii = jj = kk = ijk = -1$.
Conceptually, the jump from complex numbers to quaternions is like the jump from real numbers to complex numbers~\cite{Dickson19}.
Moreover, in the same way that for $n$-dimension vector-space the orthogonal group $\Or{n}$ preserves an inner product between real vectors, and the unitary group $\Un{n}$ preserves an inner product between complex vectors, there is a group known as the {\em symplectic group} $\Sp{n}$ which preserves an inner product for quaternionic vectors~\cite{Baez14}.

Before we can talk about the quaternionic phase-group, we remark that it is indeed meaningful to talk about quaternionic measurements.
Much like in quantum theory, one can find positive semi-definite quaternionic matrices $\{E_i\}$ which act as effects that assign probabilities to on quaternionic states $\rho$ according to $P = \tr\left(E_i \rho \right)$, and constitute a quaternionic measurement (guaranteed to give one outcome) when $\sum_i E_i = \id$.
Moreover one can find sets of $N$ such matrices satisfying $E_i E_j = \delta_{ij} E_i$ which constitute orthonormal projective measurements~\cite{Graydon11}.
Much like the trace on complex matrices, the quaternionic trace is basis-independent, and its arguments may be cyclically rotated~\cite{Graydon11}.

\begin{lemma}
For an $N$-level quaternionic quantum system, with a measurement $Z$ which distinguishes perfectly between $N$ possibilities,
 the group of diagonal quaternionic matrices
\begin{equation}
\bigoplus^N \Sp{1},
\end{equation}
are within the group of reversible phase operations (i.e.\ satisfying \cref{def:phase}).
\begin{proof}

Take a set of quaternionic projectors $\{Z_i\}$, and consider the most general reversible transformation $S \in \Sp{d}$, which acts on a quaternionic state $\rho$ as $\rho\to S \rho S^\ddag$ (we shall use $\ddag$ to represent the symplectic inverse such that $S S^\ddag = S^\ddag S = \id$).
The phase group condition in \cref{def:phase} may be written in terms of quaternionic states, effects and transformations as:
\begin{equation}
\label{eq:QPhaseTransForm}
\tr\left(Z_j  S \rho S^\ddag \right) = \tr\left(Z_j \rho \right) \qquad \forall \rho, j.
\end{equation}

If $[Z_j, S]=0$ for all $j$, then this will automatically be true for all $\rho$.
However, we should be careful, as unlike complex matrices, quaternionic matrices do not necessarily commute when they are diagonal (because scalar quaternions do not in general commute), and so we cannot automatically conclude that diagonal quaternionic matrices will satisfy this requirement.
However, we can use the quaternionic spectral theorem to find a basis where all $Z_j$ are real and diagonal.
In this basis the diagonality of $S$ is sufficient to ensure that it can commute with $Z_i$ (because $ah = ha$ for $a\in \mathcal{R}$, $h\in\quat$).
Hence, we can choose an diagonal matrix of quaternions for $S$, which in order to satisfy $S^\ddag S = \id$ must be of the form $\bigoplus \Sp{1}$ (i.e.\ each diagonal element is a unit quaternion).
\end{proof}
\end{lemma}

For the discussion in this text, we do not need to claim this is the most general quaternionic phase operation (although it may be so).
However, we will show that within this particular group, we can still find non-trivial subgroups of branch local operations.

\begin{lemma}
An operation that induces an element of $\Sp{1}$ on a branch $j$, and does not change the other branches except by a global phase of $Z_2$ is consistent with branch locality as applied to branch $j$.
\begin{proof}
The proof is similar to the case with complex quantum theory.
For $S$ to be local to branch $j$ with associated measurement effect $Z_j$, then any state where $\tr\left(Z_j \rho\right) = 0$ must not be changed.
For simplicity, we shall use a basis where $Z_j$ is real and diagonal (as is allowed by the spectral theorem).

The density matrix of states with no support in branch $j$ has element $\rho_{jj}=0$.
We can argue that this means all other elements in row $j$ and column $j$ must be $0$, using similar logic as holds in the usual quantum case:
Pure states in quaternionic quantum theory are rank-1 projectors (Lemma 3.1.1 in \cite{Graydon11}), which may be written $\ketbra{\phi}{\phi}$ (for quaternionic bras and kets); and so any pure state with no support in branch $j$ will have $\phi_j = 0$; and as $0$ conjugates to itself, this means that the entire row $j$ and column $j$ in the matrix are $0$.
As mercifully $h h^\ddag\geq 0$ for $h \in \quat$, if a state has no support for branch $j$, then every pure state it is composed from (which is always possible according to the spectral theorem) must have no support for branch $j$. 
Each of these will also only have $0$s in row $j$ and column $j$, and so all of these elements in $\rho$ will be $0$.

When $S$ is a diagonal matrix where every diagonal element is $1$ except for $S_{jj}$, branch locality will therefore automatically be respected for density matrices with $0$s in row $j$ and column $j$.
Hence, we can associate this choice of $S\in\Sp{1}$ with that branch.

Unlike with quantum theory, the global phase freedom for quaternions cannot be the same $\Sp{1}$ phase freedom available to each branch:
for $h \in \Sp{1}$, application of the global phase $G = \diag\left(h, \ldots h\right)$ gives $G \rho G^\ddag = h \rho h^\ddag$, which amounts to conjugating every element in $\rho$ with $h$. 
Because quaternions do not commute, in general this will have a non-trivial effect on $\rho$ (even if $h$ is restricted to a complex number).
Only the real elements of $\Sp{1}$, which correspond to $\{1, -1\} = Z_2$, commute and thus are never observable globally.
Hence only $Z_2$ can be taken as global phase operations which have no effect on any observable state (see also Lemma 33 and discussion in Appendix B of \cite{BarnumMU14}).
\end{proof}
\end{lemma}

This works to our advantage: unlike a complex global phase, a quaternionic global phase would cause us problems with commutativity: if two spatially disjoint agents could both act to span the entire phase-group of a quaternionic two-level system, then relativistic observers might disagree on what the resultant transformation should be, and this would ultimately conflict with the objective reality associated with probabilistic clicks in the detector~\cite{GarnerMD14}.
Without a quaternionic global phase, branch locality can divide the quaternionic phase group up into subgroups, each of which with elements that commute with all elements in the other subgroups.
(The $N=2$ case is discussed in \cite{GarnerMD14}, where the two phase groups end up corresponding to the left and right isoclinic rotations in $\SO{4}$, and the only shared element other than identity corresponds to the inversion $1\to-1$, which is what we require for our interferometric computation.)

\end{document}